\documentclass[nofootinbib,10pt,twocolumn,preprintnumbers]{revtex4-1}
\usepackage{amsmath,amssymb,pgf,pgfarrows,pgfnodes,float,appendix, hyperref,slashed,breakurl,graphicx}
\definecolor{Color}{rgb}{0.28, 0.24, 0.55}
\definecolor{Orange}{rgb}{1,0.38,0.11}
\hypersetup{
    colorlinks = true,
    citecolor = Orange,
    linkcolor = Color,
    urlcolor  = Orange,
}
\definecolor{internationalorange}{rgb}{1.0, 0.31, 0.0}

\usepackage{graphicx}
\usepackage{subfigure}
\usepackage[margin=0.9in]{geometry}

\usepackage{enumerate}

\usepackage{amsmath}

\usepackage{color, colortbl}
\definecolor{Gray}{gray}{0.8}
\definecolor{GrayLight}{gray}{0.4}
\definecolor{Darkgreen}{RGB}{30,120,30}
\definecolor{granate}{rgb}{0.8039,0.2,0.2}
\newcommand{\beq}{\begin{equation}}
\newcommand{\eeq}{\end{equation}}
\newcommand{\bea}{\begin{eqnarray}}
\newcommand{\eea}{\end{eqnarray}}

\usepackage{eso-pic}

\usepackage{xparse}

\usepackage{tikz}
\usetikzlibrary{"arrows", "automata", "backgrounds", "calendar", "chains", "matrix", "mindmap", "patterns", "petri", "shadows", "shapes.geometric", "shapes.misc", "spy", "trees"}
\usetikzlibrary{arrows,shapes}
\usetikzlibrary{trees}
\usetikzlibrary{matrix,arrows} 				
\usetikzlibrary{positioning}				
\usetikzlibrary{calc,through}				
\usetikzlibrary{decorations.pathreplacing}  
\usepackage{pgffor}							

\usetikzlibrary{decorations.pathmorphing}	
\usetikzlibrary{decorations.markings}
\tikzset{
    vector/.style={decorate, decoration={snake}, draw},
	provector/.style={decorate, decoration={snake,amplitude=2.5pt}, draw},
	antivector/.style={decorate, decoration={snake,amplitude=-2.5pt}, draw},
    fermion/.style={draw=black, postaction={decorate},
        decoration={markings,mark=at position .55 with {\arrow[draw=black]{>}}}},
    fermionr/.style={draw=black, postaction={decorate},
    decoration={markings,mark=at position .55 with {\arrow[draw=black]{<}}}},
    fermioncyan/.style={draw=black, postaction={decorate},
        decoration={markings,mark=at position .55 with {\arrow[draw=cyan]{<}}}},
    fermiondif/.style={draw=black, postaction={decorate},
        decoration={markings,mark=at position .7 with {\arrow[draw=black]{>}}}},
            fermiondif2/.style={draw=black, postaction={decorate},
        decoration={markings,mark=at position .7 with {\arrow[draw=black]{<}}}},
    fermionend/.style={draw=black, postaction={decorate},
        decoration={markings,mark=at position 1 with {\arrow[draw=black]{>}}}},
    fermionuchannel2/.style={draw=black, postaction={decorate},
        decoration={markings,mark=at position .4 with {\arrow[draw=black]{>}}}},
    scalardif/.style={dashed,draw=black, postaction={decorate},
        decoration={markings,mark=at position .7 with {\arrow[draw=black]{>}}}},
    scalarend/.style={dashed,draw=black, postaction={decorate},
        decoration={markings,mark=at position 1 with {\arrow[draw=black]{>}}}},
    fermionbar/.style={draw=black, postaction={decorate},
        decoration={markings,mark=at position .55 with {\arrow[draw=black]{<}}}},
    fermionnoarrow/.style={draw=black},
    gluon/.style={decorate, draw=black,
        decoration={coil,amplitude=4pt, segment length=5pt}},
    scalar/.style={dashed,draw=black, postaction={decorate},
        decoration={markings,mark=at position .55 with {\arrow[draw=black]{>}}}},
    scalarcyan/.style={dashed,draw=black, postaction={decorate},
        decoration={markings,mark=at position .55 with {\arrow[draw=cyan]{>}}}},
    scalaruchannel1/.style={dashed,draw=black, postaction={decorate},
        decoration={markings,mark=at position .7 with {\arrow[draw=black]{>}}}},
                  scalaruchannel2/.style={dashed,draw=black, postaction={decorate},
        decoration={markings,mark=at position .4 with {\arrow[draw=black]{>}}}},
    scalarbar/.style={dashed,draw=black, postaction={decorate},
        decoration={markings,mark=at position .55 with {\arrow[draw=black]{<}}}},
    scalarnoarrow/.style={dashed,draw=black},
    electron/.style={draw=black, postaction={decorate},
        decoration={markings,mark=at position .55 with {\arrow[draw=black]{>}}}},
	bigvector/.style={decorate, decoration={snake,amplitude=4pt}, draw},
}

\NewDocumentCommand\semiloop{O{black}mmmO{}O{above}}
{%
\draw[#1] let \p1 = ($(#3)-(#2)$) in (#3) arc (#4:({#4+180}):({0.5*veclen(\x1,\y1)})node[midway, #6] {#5};)
}

\tikzstyle{block} = [draw, rectangle, 
    minimum height=3em, minimum width=6em]

\usepackage{enumitem}
\usepackage{tikz}
\usetikzlibrary{fit}
\tikzset{%
  highlight/.style={rectangle,rounded corners,color=granate,draw,text opacity =1,
    fill opacity=0.5,thick,inner sep=0pt}
}

\usepackage{xparse}
\NewDocumentCommand\loopv{O{black}mmmO{}O{above}}
{%
\draw[#1] let \p1 = ($(#3)-(#2)$) in (#3) arc (#4:({#4+360}):({0.5*veclen(\x1,\y1)})node[midway, #6] {#5};)
}

\usepackage{placeins}

\tikzset{
    cross/.pic = {
    \draw[rotate = 45] (-#1,0) -- (#1,0);
    \draw[rotate = 45] (0,-#1) -- (0, #1);
    }
}

\tikzset{
    square/.style={%
        draw=none,
        circle,
        append after command={%
            \pgfextra \draw[#1] (\tikzlastnode.north-|\tikzlastnode.west) rectangle 
                (\tikzlastnode.south-|\tikzlastnode.east);\endpgfextra}
    },
    square/.default=black
}

\tikzstyle{block} = [draw, rectangle, 
    minimum height=3em, minimum width=6em]

\usepackage{xparse}

\begin{document}

\title{\Large{On Lepton and Baryon Numbers as Local Gauge Symmetries}}
\author{Pavel Fileviez P\'erez}
\affiliation{
Physics Department and Center for Education and Research in Cosmology and Astrophysics (CERCA), Case Western Reserve University, Cleveland, OH 44106, USA}

\begin{abstract} 
A simple theory where the total lepton number is a local gauge symmetry is proposed. In this context, the gauge anomalies are cancelled with the minimal number of extra fermionic fields and one predicts that the neutrinos are Majorana fermions. The properties of the neutrino sector are discussed showing that this theory predicts a $3+2$ light neutrino sector.
We show that using the same fermionic fields one can gauge the baryon number and define a simple theory where the lepton and baryon numbers can be spontaneously broken at the low scale in agreement with experiments. 
\end{abstract}

\maketitle
\section{Introduction}
The Standard Model (SM) of particle physics is one of the most successful theories of nature, describing the properties of quarks and leptons and how they interact through the strong, weak and electromagnetic interactions.
After the SM Brout-Englert-Higgs boson discovery at the LHC we know how the electroweak symmetry is broken in nature. In the SM the massive gauge bosons, the $Z^0$ and $W^\pm$, and charged fermions acquire their masses through the Higgs mechanism. Unfortunately, the SM does not provide a mechanism for neutrino masses. This is one of the main reasons why we need to modify the SM. 

There are several theories for physics beyond the SM where one can understand the origin of neutrino masses. Since the SM neutrinos could be Majorana or Dirac fermions, one needs to understand different possibilities to generate their masses. One of the most popular ideas to generate Majorana neutrino masses is based on the so-called seesaw mechanism~\cite{Minkowski:1977sc,Gell-Mann:1979vob,Glashow:1979nm,Yanagida:1979as,Mohapatra:1979ia}, where the SM neutrino masses are suppressed by a mass scale related to new physics.
The neutrinos are predicted to be Majorana fermions in Pati-Salam~\cite{Pati:1974yy} and grand unified theories~\cite{Georgi:1974my,Fritzsch:1974nn}, where quarks and leptons live in the same multiplet. Unfortunately, these theories can describe physics at the very high energy scale, $M_{GUT} \sim 10^{15-16}$ GeV, and one cannot hope to test directly the origin of neutrino masses at collider experiments.

The key symmetry one needs to study to understand the type of neutrino masses generated is the total lepton number, $\ell=\ell_e+\ell_\mu+\ell_\tau$, with $\ell_i$ being the lepton number for each SM fermionic family. The Dirac mass terms conserve the total lepton number, while the Majorana terms break the total lepton number by two units. The total lepton number in the SM is an accidental global symmetry broken at the quantum level by $SU(2)_L$ instantons. In physics beyond the SM one can have theories where $\ell$ is explicitly broken, for example in grand unified theories, or one can have theories where the local gauge symmetry $U(1)_\ell$ is spontaneously broken.

Theories based on local $U(1)_\ell$ gauge symmetry have been studied in detail in Refs.~\cite{FileviezPerez:2011pt,Duerr:2013dza,FileviezPerez:2014lnj}. 
For an earlier discussion see Ref.~\cite{Foot:1989ts}.
These theories predict a new sector needed for anomaly cancellation, new sources of CP-violation, a dark matter candidate from anomaly cancellation, and the cosmological bounds on the dark matter relic density implies that the symmetry breaking scale must be below the multi-TeV scale. In these theories the neutrinos are predicted to be Dirac fermions because the total lepton number is broken by three units.
Recently, we have investigated the implementation of the canonical seesaw mechanism in these theories~\cite{Debnath:2023akj,Debnath:2024vpf}. For several phenomenological and cosmological studies in these theories see Refs.~\cite{Schwaller:2013hqa,FileviezPerez:2019cyn,Carena:2022qpf,FileviezPerez:2015mlm,Madge:2018gfl}.

In this article, we propose a simple theory where the total lepton number is a local gauge symmetry. In this context, the gauge anomalies are cancelled with the minimal number of extra fermionic fields and one predicts that the neutrinos are Majorana fermions. We discuss the properties of the neutrino sector, showing that this theory predicts a $3+2$ light neutrino sector.
We show that using the same fermionic fields one can gauge the baryon number and define a simple theory where the lepton and baryon numbers can be spontaneously broken at the low scale because the proton is stable and other baryon number violating processes are highly suppressed.  

This article is organized as follows: In Section~\ref{sec2} we discuss the leptonic anomalies and the new solution for anomaly cancellation. We discuss the neutrino spectrum and point out the existence of two light right-handed neutrinos. In Section~\ref{sec3} we discuss the possibility to gauge the baryon number using the same fermionic representations used in Section~\ref{sec2} and define a simple theory for spontaneous baryon and lepton numbers. We summarize our main results in Section~\ref{summary}.
\section{Local Lepton Number}
\label{sec2}
In order to define a gauge theory for total lepton number and understand the spontaneous symmetry breaking, one needs to make sure that the theory is anomaly free. In this context, using the SM leptonic fields, 
\begin{displaymath}
\ell_L \sim ({\bf{2}},-1/2,1) \ {\rm{and}} \  e_R \sim ({\bf{1}},-1,1),
\end{displaymath}
one can estimate the different gauge anomalies:
\begin{eqnarray*}
{\mathcal{A}}_1(SU(3)_C^2 U(1)_\ell) &=&0, \\
{\mathcal{A}}_2(SU(2)_L^2 U(1)_\ell) &=&3/2, \\
{\mathcal{A}}_3(U(1)_Y^2 U(1)_\ell) &=& -3/2, \\
{\mathcal{A}}_4(U(1)_Y U(1)_\ell^2) &=& 0,\\
{\mathcal{A}}_5(U(1)_\ell^3)=3,  &\text{and}& 
{\mathcal{A}}_6 (U(1)_\ell)=3.
\end{eqnarray*}
Here we show only how the leptonic fields transform under the $SU(2)_L$, $U(1)_Y$ and $U(1)_\ell$ gauge groups.
The last two anomalies, ${\mathcal{A}}_5(U(1)_\ell^3)$ and ${\mathcal{A}}_6 (U(1)_\ell)$, can be cancelled if one adds three copies of right-handed neutrinos, $\nu_R \sim ({\bf{1}},0,1)$, needed to implement the canonical seesaw mechanism. 

Notice that since ${\mathcal{A}}(SU(2)_L^2 U(1)_\ell) \neq 0$ one needs at least one fermionic representation transforming non-trivially under $SU(2)_L$. Different solutions have been proposed for anomaly cancellation in this context: a) Vector-like leptons~\cite{FileviezPerez:2011pt,Duerr:2013dza}, b) Four fermionic representations~\cite{FileviezPerez:2014lnj}.
If one stick to the minimal number of fields needed to understand the spontaneous breaking of total lepton number, the neutrinos are Dirac fermions in these theories because the total lepton number is broken by three units. The case of Majorana neutrinos has been also discussed in Refs.~\cite{Debnath:2023akj,Debnath:2024vpf}.

In this article, we would like to point out a simple solution for anomaly cancellation that allows us to construct a theory for spontaneous lepton number breaking. If we study the anomaly cancellation conditions, one can find a simple solution where all anomalies are cancelled with only four fermions, and only one of the extra fermionic representations is in a non-trivial representation of $SU(2)$:
\begin{eqnarray}
    \Psi_L & \sim & ({\bf{1}},-1,3/4), \
    \Psi_R  \sim  ({\bf{1}},-1,-3/4), \nonumber \\
    \chi_L & \sim & ({\bf{1}},0,3/4), \ \text{and} \ 
    \rho_L   \sim  ({\bf{3}},0,-3/4). \nonumber
    \label{fermions}
\end{eqnarray}
Here we do not show the QCD quantum numbers because all these fields are colorless, i.e. they do not transform under $SU(3)_C$. Notice that the lepton number for these new fermionic fields is predicted by anomaly cancellation. 

In Ref.~\cite{FileviezPerez:2014lnj} we pointed out a solution with only three representations but in this case one predicts fermions with fractional electric charge that cannot decay into the SM fields. Also in Ref.~\cite{FileviezPerez:2014lnj} we pointed out a solution with four fermionic fields but three of them are in non-trivial representations of $SU(2)$, and one has more physical fields after symmetry breaking if one compares to the new solution proposed in this article. 

One can generate mass for the new fermions if one has a Higgs field, $S \sim ({\bf{1}},0,3/2)$, and the following Yukawa interactions:
\begin{eqnarray}
- \mathcal{L} &\supset& \lambda_\rho {\rm{Tr}}(\rho_L^T C \rho_L) S + \lambda_\Psi \bar{\Psi}_L \Psi_R S \nonumber \\
&+& \lambda_\chi \chi_L^T C \chi_L S^* \ + \ {\rm{h.c.}}.
\label{interactions-masses}
\end{eqnarray}
Notice that here the trace in ${\rm{Tr}}(\rho_L \rho_L)$ is for the $SU(2)$ index. The $\rho_L$ field can be written as
\begin{equation}
\rho_L=\frac{1}{2} 
\left( \begin{matrix}
\rho_L^0 & \sqrt{2} \rho_L^+ \\
\sqrt{2} \rho_L^- & - \rho_L^0
\end{matrix}
\right).
\end{equation}
The new electrically charged fields, $\Psi_L$ and $\Psi_R$, can decay if they mix with the SM leptons. This can occur if we have the first interaction in the following equation:
\begin{eqnarray}
- \mathcal{L} &\supset& \lambda_e \bar{\Psi}_L e_R \phi + \lambda_R \bar{\chi}_L \nu_R \phi \nonumber \\
&+& y_\nu \bar{\ell}_L i \sigma_2 H^* \nu_R + y_e \bar{\ell}_L H e_R  \nonumber \\  
& + & \frac{y}{\Lambda} \ell_L^T i \sigma_2 C \rho_L H \phi \ + \ {\rm{h.c.}}.
\label{interactions-decays}
\end{eqnarray}
Here $\phi \sim ({\bf{1}},0,-1/4)$ and $H \sim({\bf{2}},1/2,0)$ is the SM Higgs field.
The second term in the above equation is also allowed by the gauge symmetry and it has very important consequences for the mechanism for neutrino masses. One could consider a different possibility where the $\Psi_R$ can mix with the SM leptonic doublet, $\ell_L$, if one adds an $SU(2)_L$ Higgs doublet with more exotic lepton number, $H_\ell \sim ({\bf{2}},1/2,7/2)$.
Since our main interest is to look for the simplest theory based on local $U(1)_\ell$, we focus on the theory where one uses the interactions in Eq.(\ref{interactions-decays}). Therefore, the minimal theory contains two extra scalar fields: $S \sim ({\bf{1}},0,3/2)$ and  $\phi \sim ({\bf{1}},0,-1/4)$. Their vacuum expectation values are defined as: $\left<\phi\right>=v_\phi/\sqrt{2}$ and $\left<S\right>=v_S/\sqrt{2}$.
\begin{itemize}
\item {\textit{Neutrino Masses}}:
In this theory, the SM neutrino masses are generated through a double-seesaw mechanism.
After $U(1)_\ell$ is spontaneously broken one can generate the mass matrix for the right-handed neutrinos using the terms:
\begin{equation}
    - \mathcal{L}_\nu \supset \lambda_R^i \frac{v_\phi}{\sqrt{2}} \bar{\chi}_L \nu_R^i + \lambda_\chi \frac{v_S}{\sqrt{2}} \chi_L^T C \chi_L  \ + \ \rm{h.c.}, 
\end{equation}
and integrating out the $\chi_L$ field. The mass matrix for the right-handed neutrinos reads as
\begin{equation}
M_{\nu_R}^{ij} = \frac{\lambda_R^i \lambda_R^j v_\phi^2}{2 \sqrt{2} \lambda_\chi v_S}. 
\end{equation}
Here $i,j=1,2,3$.
Notice that this mass matrix has rank one. Therefore, only one right-handed neutrino is massive after $U(1)_\ell$ is broken. One can rotate the right-handed neutrinos, $\nu_R \to N_R \nu_R$, and write the mass terms for the SM neutrinos and right-handed neutrinos in the following way:
\begin{eqnarray}
- \mathcal{L}_{\nu}^m &\supset& \bar{\nu}_L^i (\tilde{M}_D^{\nu})^{ij} \nu_R^j - \frac{1}{2} \nu_R^{3 T} M_R \ C \nu_R^3 \nonumber \\
&+& \frac{M_\rho}{2} (\rho_L^0)^T  C \rho_L^0 
- \frac{y v_\phi v_0}{4 \Lambda} \nu_L^T C \rho_L^0 \ + \ \rm{h.c.}, \nonumber \\
\end{eqnarray}
with $\tilde{M}_D^{\nu}=y_\nu v_0 N_R/\sqrt{2}$ and $M_\rho=\lambda_\rho v_S/\sqrt{2}$. Here $v_0=246$ GeV is the vacuum expectation value of the SM Higgs field and $M_R$ is the mass of the massive right-handed neutrino. Here we use the convention where only the third right-handed neutrino field is massive. In the above equation we included the mass for $\rho_L^0$ and the mixing term between $\nu_L$ and $\rho_L^0$.

Integrating out the massive right-handed neutrino and $\rho_L^0$ one finds:
\begin{equation}
- \mathcal{L}_{\nu}^m \supset \bar{\nu}_L^i (\tilde{M}_D^{\nu})^{i\alpha} \nu_R^\alpha - \frac{1}{2} m_\nu^{ij} \nu_L^{i T} C \nu_L^j \ + \ \rm{h.c.}.
\label{numasses}
\end{equation}
Here $\alpha=4,5$ and 
\begin{equation}
m_\nu^{ij}= \frac{(\tilde{M}_D^{\nu})^{i 3}(\tilde{M}_D^{\nu})^{j 3}}{M_R} + \frac{m_D^i m_D^j}{M_\rho}.
\end{equation}
Here $m_D^i=y^i v_0 v_\phi/4 \Lambda$. Notice that $m_\nu^{ij}$ has rank two. Therefore, this theory predicts a 3+2 light neutrino sector. This theory does not predict the numerical values of the coefficients in the above mass matrix, but one expects that the two extra sterile neutrinos can have masses smaller than the SM neutrino masses with small mixing angles. This is possible when $(\tilde{M}_D^{\nu})^{i\alpha}$ is very small. Notice that since $m_\nu^{ij}$ has rank two, one of the SM neutrinos must be massless. We will study in detail the cosmological bounds in this model in a future publication.

In this discussion, we did not include the dimension five operators such as $H^\dagger \rho_L \chi_L H / \Lambda$ because generically it provides a small corrections to the $\rho_L$ and $\chi_L$ masses. One could have the dimension six operator: $c_R \nu_R \nu_R \phi^2 S^*/\Lambda^2$ that can change the predictions for the right-handed neutrino masses. Assuming $c_R \sim 1$, $v_\phi \sim v_S \sim 1$ TeV and $\Lambda > 10^5$ TeV (see discussion about proton decay), one finds that $\delta M_R <  50$ eV. Notice that if this operator is present one changes the tree-level predictions for the right-handed neutrino masses above, but still two right-handed neutrinos are very light.  

\item {\textit{Higgs Sector}}:
The scalar potential in this theory is given by 
 \begin{eqnarray}
 V(H,S,\phi)&=&-m_H^2 H ^{\dagger}H+\lambda(H^{\dagger}H)^2-m_s^2 S ^{\dagger}S \nonumber \\
 &+& \lambda_s (S^{\dagger}S)^2-m_{\phi}^2 \phi ^{\dagger}\phi 
 + \lambda_{\phi}(\phi^{\dagger}\phi)^2 \nonumber \\
 &+& \lambda_1(H^{\dagger}H)S^{\dagger}S + \lambda_2(H^{\dagger}H)\phi^{\dagger}\phi \nonumber \\
 &+& \lambda_3(S^{\dagger}S)\phi^{\dagger}\phi.
 \end{eqnarray}
The scalar fields can be written as 
\begin{eqnarray}
H&=&\begin{pmatrix}
h^+\\
\frac{1}{\sqrt{2}}(v_{0} + h_0) e^{i \sigma_0/v_0} 
\end{pmatrix}, \\
S &=& \frac{1}{\sqrt{2}}\left(v_{s} + h_s \right) e^{i \sigma_s/v_s}, 
\label{S-filed}
\end{eqnarray}
and
\begin{eqnarray}
\phi &=& \frac{1}{\sqrt{2}}\left( v_{\phi} + h_{\phi} \right) e^{i \sigma_{\phi}/v_{\phi}} \label{phi-field}.
\end{eqnarray}
Notice that this scalar potential has the global symmetry: 
$O(4)_H \otimes U(1)_\phi \otimes U(1)_S $. The physical CP-even Higgses, $(h,H_1,H_2)$, are defined as
\begin{eqnarray}
\begin{pmatrix}
h_0\\
h_s\\
h_{\phi} 
\end{pmatrix} =
U
\begin{pmatrix}
h\\
H_1\\
H_2
\end{pmatrix}.
\label{Umixing}
\end{eqnarray}
There are three CP-odd Higgses, two of them are Goldstone's bosons eaten by the neutral gauge bosons. $U(1)_\ell$ allows a dimension seven term in the potential: $\lambda_{M} S \phi^6/\Lambda^3 + {\rm h.c.},$ which breaks the $U(1)_\phi \otimes U(1)_S$ symmetry of the potential and one gets a pseudo-Nambu-Goldstone boson, the Majoron $J$. In the next section, we will discuss the possibility to gauge also baryon number and in this case all the CP-odd fields are eaten by the neutral gauge bosons. The dimension seven operator mentioned above is not allowed by the $U(1)_B$ gauge symmetry as we will discuss.

\item {\textit{Charged Leptons}}: The SM charged leptons mix with the new singly charged fermions in the theory through the interactions in Eq.(\ref{interactions-decays}). The mass matrix for the charged leptons in the basis, $(e_L, \Psi_L)$ and $(e_R, \Psi_R)$, is given by
\begin{equation}
{\mathcal{M}}_c=\left(\begin{matrix}
y_e v_0/\sqrt{2} & 0 \\
\lambda_e v_\phi/\sqrt{2} & \lambda_\Psi v_S/\sqrt{2} 
\end{matrix} \right).
\end{equation}
The mixing between the SM charged leptons and the new charged lepton is very small since $v_\phi, v_S \gg v_0$. This simple theory predicts lepton flavour violation mediated by the interactions in Eq.(\ref{interactions-decays}) that can give rise to lepton flavour violating processes such as $\mu \to e \gamma$. We will discuss the predictions for these processes in a future publication. The charged components in $\rho_L$ mix with the SM charged leptons via the dimension five operator in Eq.(\ref{interactions-decays}). 

\item {\textit{Proton Decay:}} In this theory proton decay can be mediated by the dimension nine operators\footnote{Thanks to 
C. Murgui and the referee for this comment.}. For example, one can have these operators:
\begin{equation}
\frac{c_\ell}{\Lambda^6} q_L q_L q_L \ell_L S^* (\phi^*)^2 + \frac{c_e}{\Lambda^6} u_R u_R d_R e_R S^* (\phi^*)^2.
\end{equation}
Using $c_\ell \sim 1$, $c_e \sim 1$, $v_\phi \sim v_S \sim 1$ TeV, one finds that $\Lambda > 10^5$ TeV in order to satisfy the experimental bounds on the proton decay lifetime. For a review on proton decay see Ref.~\cite{Nath:2006ut}. Notice that these operators are not allowed in the full theory that will be discuss in the next section, where the baryon number is also gauged.

\end{itemize}
In this section, we have assumed the scalar field $\phi$ acquires a vacuum expectation value generating the mixings between the new fields and the SM leptons. This allows us to generate the neutrino masses and as we will discuss in the next section, we need only two scalar fields, $S$ and $\phi$, to define a realistic theory where the total lepton and baryon numbers are spontaneously broken. If $\phi$ does not have a vacuum expectation value, the model can be realistic because the new fields can decay into the SM fields and one of the new neutral fields. In this case, the lightest field between $\phi$, $\chi_L$ and $\rho_L^0$ can be a dark matter candidate.
\section{Local Baryon Number}
\label{sec3}
It has been shown in previous studies that in a theory with local lepton number one can gauge the baryon number with the same number of extra fermions~\cite{Duerr:2013dza,FileviezPerez:2014lnj}. Therefore, one can have a simple gauge theory based on 
\begin{eqnarray}
SU(3)_C \otimes SU(2)_L \otimes U(1)_Y \otimes U(1)_\ell \otimes U(1)_B.  \nonumber
\end{eqnarray}
Using the SM quark fields:
$q_L \sim ({\bf{3}},{\bf{2}},1/6,0,1/3)$, $u_R \sim ({\bf{3}},{\bf{1}},2/3,0,1/3)$
and $d_R \sim ({\bf{3}},{\bf{1}},-1/3,0,1/3)$, one can estimate the baryonic gauge anomalies:
\begin{eqnarray*}
 {\mathcal{A}}_7(SU(3)_C^2 U(1)_B)&=&0, \\
{\mathcal{A}}_8(SU(2)_L^2 U(1)_B)&=&3/2, \\
{\mathcal{A}}_9(U(1)_Y^2 U(1)_B)&=&-3/2, \\
{\mathcal{A}}_{10}(U(1)_Y U(1)_B^2)&=&0, \\
{\mathcal{A}}_{11}(U(1)_B^3)=0, \ &\text{and}&  \
{\mathcal{A}}_{12} (U(1)_B)=0.
\end{eqnarray*}
In previous studies three simple ways to cancel the baryonic anomalies have been discussed:
a) adding six extra colorless fermionic representations~\cite{Duerr:2013dza}, b) adding four extra colorless fermionic representations~\cite{FileviezPerez:2014lnj} or
c) adding vector-like quarks~\cite{FileviezPerez:2011pt}. In the previous section we listed the needed fields to cancel the leptonic anomalies. Using the same fields
we can cancel all baryonic anomalies if these extra fermions have also baryon number and with the following quantum numbers:
\begin{eqnarray}
    \Psi_L & \sim & ({\bf{1}}, {\bf{1}},-1,3/4,3/4), \nonumber \\
    \Psi_R  & \sim & ({\bf{1}},{\bf{1}},-1,-3/4,-3/4), \nonumber \\
    \chi_L & \sim & ({\bf{1}},{\bf{1}},0,3/4,3/4), \nonumber \\
    \rho_L  & \sim & ({\bf{1}},{\bf{3}},0,-3/4,-3/4). \nonumber
    \label{fermions}
\end{eqnarray}
Notice that one can also cancel all mix anomalies:
\begin{eqnarray}
 && {\mathcal{A}}_{13}(U(1)_B^2 U(1)_\ell), \
{\mathcal{A}}_{14}(U(1)_B U(1)_\ell^2), \nonumber \\
&& \ \text{and} \ {\mathcal{A}}_{15}(U(1)_Y U(1)_\ell U(1)_B). \nonumber 
\end{eqnarray}
We refer to these extra fields as {\textit{lepto-baryons}} because they have lepton and baryon numbers. 

The new scalar fields needed for symmetry breaking and mass generation are the same as in the previous section but now we need to define the corresponding baryon numbers:
\begin{eqnarray}
S & \sim & ({\bf{1}}, {\bf{1}},0,3/2,3/2), \ \text{and} \
\phi \sim ({\bf{1}}, {\bf{1}},0,-1/4, 3/4). \nonumber 
\end{eqnarray}
Notice that in this theory the two extra Goldstone's bosons, $\sigma_s$ and $\sigma_\phi$, in Eqs. (\ref{S-filed}) and (\ref{phi-field}) are eaten by the two new neutral gauge bosons associated to $U(1)_B$ and $U(1)_\ell$. 
\begin{itemize}

\item {\textit{New Gauge Bosons}}: In this theory we have two new neutral gauge bosons, $Z_\ell$ and $Z_B$, associated to $U(1)_\ell$ and $U(1)_B$, respectively. The mass matrix for these gauge bosons in the basis, $(Z_\ell^\mu, Z_B^\mu)$, can be written as
\begin{equation}
   {\mathcal{M}}^2_0=\left(\begin{matrix}
         g_\ell^2 (\frac{9}{4} v_S^2 + \frac{1}{16} v_\phi^2)& g_\ell g_B (\frac{9}{4} v_S^2 - \frac{3}{16} v_\phi^2)\\ \\
         g_\ell g_B (\frac{9}{4} v_S^2 - \frac{3}{16} v_\phi^2) & g_B^2 (\frac{9}{4} v_S^2 + \frac{9}{16} v_\phi^2)
    \end{matrix} \right).
\end{equation}
Here we neglect the kinetic mixing between the Abelian gauge bosons for simplicity. Notice that the mixing between the gauge bosons can be large. The SM quarks couple to $Z_B$, while the SM leptons couple to $Z_\ell$. However, since the mixing angle between the new neutral gauge bosons can be large, the new physical neutral gauge bosons couple to the SM quarks and leptons. For experimental limits on the new neutral gauge boson masses that couple simultaneously to quarks and leptons see Ref.~\cite{Alioli:2017nzr}.

\item {\textit{Baryon Number Violating Processes}}: In this theory the operators mediating proton decay are not allowed because baryon number is not broken by one unit. Therefore, this theory predicts that the proton is stable. In this theory, one can have for example the following operator
\begin{equation}
    \frac{c_{BL}}{\Lambda^{16}} (q_L q_L q_L \ell_L)^3 (S^*)^2, \\
\end{equation}
which mediates the processes: $ppp \to e^+ e^+ e^+$, $ppn \to e^+ e^+ \bar{\nu}$, $pnn \to e^+ \bar{\nu} \bar{\nu}$ and $3n \to 3 \bar{\nu}$. However, these processes are highly suppressed even if the cut-off scale $\Lambda \sim 1$ TeV due to the fact that these operators are suppressed by $\Lambda^{16}$. Since in this theory the proton is absolutely stable and the other possible baryon violating processes are highly suppressed, the symmetry breaking scale for $U(1)_B$ can be close to the electroweak scale.
\end{itemize}

In this section, we have discussed the full theory where the baryon and total lepton numbers are local gauge symmetries. One can consider the case where only the baryon number is a local gauge symmetry and cancel all the baryonic anomalies with the four extra fermions, $\Psi_L,\Psi_R,\rho_L$ and $\chi_L$, listed above. In this case, the $\phi$ scalar field is still needed to allow the new $\Psi$-fields to decay. Now, in the scenario where $\phi$ does not acquire a vacuum expectation value, the lightest field between $\phi$, $\chi_L$ and $\rho_L^0$ is stable and it can be a cold dark matter candidate. In this case, one has an accidental discrete symmetry protecting the stability of the lightest field, i.e. $\mathcal{Z}_2$: $\phi \to - \phi$, $\chi_L \to - \chi_L$, $\rho_L \to - \rho_L$, $\Psi_L \to - \Psi_L$, and $\Psi_R \to - \Psi_R$. This theory will be investigated in a future publication. 

\section{Summary}
\label{summary}
In this article, we have discussed the simplest theory based on local total lepton number where the gauge anomalies are cancelled with only four fermions plus the right-handed neutrinos needed for the seesaw mechanism. The neutrinos are predicted to be Majorana fermions.
The neutrino masses are generated through a double seesaw mechanism predicting a light $3+2$ neutrino sector, where one expects that the two extra sterile neutrinos can have mass below or at the eV scale.
We have shown that using the same extra fermionic fields one can cancel all baryonic anomalies and define the simplest theory where the total lepton and baryon numbers can be spontaneously broken at the low scale. In this theory, the proton is stable and other baryon number violating processes are highly suppressed. We also briefly discussed the scenarios where only one symmetry, baryon or total lepton number, is gauged. We discussed the possibility to have a dark matter candidate in these scenarios.

{\small{\textit{Acknowledgments:}}}
I would like to thank C. Murgui and M. B. Wise for discussions. This work has been supported by the U.S. Department of Energy, Office of Science, Office of High Energy Physics, under Award Number DE-SC0024160. I would like to thank the referees for very useful comments.

\bibliography{refs}

\end{document}